\documentclass{aa}

\usepackage{graphics}
\begin{document}

   \thesaurus{01
              (13.07.1;  
               )}
   \title{Evidences for two Gamma-Ray Burst\\afterglow
   emission regimes}

   \subtitle{}

   \titlerunning{GRB Afterglow emission regimes}

   \author{M. Bo\"er \and B. Gendre
          }

   \offprints{M. Bo\"er}

   \institute{Centre d'Etude Spatiale des Rayonnements (CNRS/UPS), BP 4346,
   31028 Toulouse Cedex, France,\\
              email: Michel.Boer@cesr.fr
             }

   \date{Received 17 July 2000; Accepted 24 August 2000}

   \maketitle

   \begin{abstract}

   We applied cosmological and absorption corrections
   to the X-ray and optical afterglow fluxes
   of a sample of Gamma-Ray Burst sources of known distance.
   We find a good correlation in X-rays
   and that the GRBs in our sample form
   two well defined classes. We tentatively interpret them as
   radiative and adiabatic afterglow behaviours in the framework of the
   fireball model for GRBs. We do not observe this correlation
   at optical wavelengths.
   This discrepancy with the model
   may be due to the absorption in the source
   vicinity.


   \end{abstract}

%

\section{Introduction}

   The detection of X-ray and optical afterglows of
   cosmic gamma-ray bursts (hereafter GRB) has firmly established
   the fireball model (Rees and M\'esz\'aros \cite{Rees92},
   M\'esz\'aros and Rees \cite{Mesz97}, Panaitescu
   et al. \cite{Pana98}) as a standard tool to interpret GRB afterglows. A
    vast majority of  sources have X-ray afterglows,
while about half of them have been observed at optical
wavelengths. The fireball model provides firm predictions on the
 temporal behaviour of the afterglow emission at all wavelengths,
  allowing an inter-comparison of different events (Piran
  \cite{Piran99}, Sari et al. \cite{Sari98}).
  In this framework the afterglow emission is described as synchrotron emission of
accelerated electrons during the shock of an ultra-relativistic
shell with the external medium.

In this letter, we use a set of GRB sources detected during their
afterglow both at X-ray and optical wavelengths, and for which a
firm measure of the source distance has been established. We
apply several distance corrections and we take into account the
galactic absorption for the optical data. The light-curve of each
source of our sample is computed for a standard distance
corresponding to a redshift of 1.

\section{The burst sample}

\begin{figure*}
\resizebox{\hsize}{!}{\includegraphics{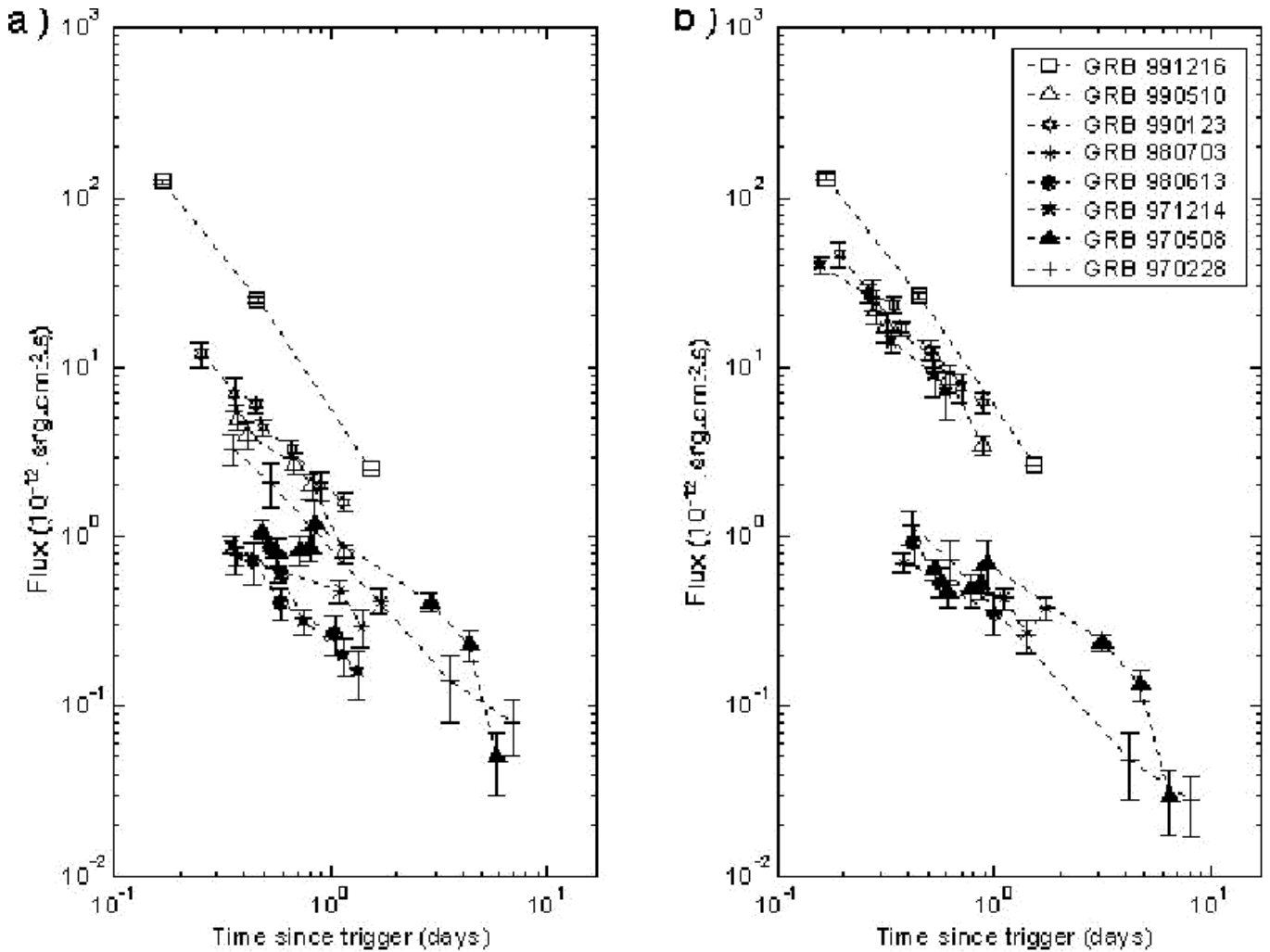}} \caption{X-ray
light curves of the burst sample a) no correction has been
applied, and b) with distance, flux, and time dilation
normalization to a redshift of 1 applied} \label{XFig}
\end{figure*}

We used a sample of eight GRBs detected both at X-ray, by the NFI
instrument on board the BeppoSAX satellite, and optical
wavelengths. A redshift measure is available for each of the
sources of our sample. We did not use the data from GRB 980425
since its association with SN 1998bw remains questionable. The
redshifts and the main references on the sources  of our sample
have been summarized on table 1.

\begin{table}
\caption[]{The GRB source sample}
\label{dataused}
\[
         \begin{array}{lll}
            \hline
            \noalign{\smallskip}
           Source&Redshift&References\\
            \noalign{\smallskip}
            \hline
\object{GRB\,970228} & 0.695 & 1, 2 \\
\object{GRB\,970508} & 0.835 & 1,3  \\
\object{GRB\,971214} & 3.42 & 1, 4  \\
\object{GRB\,980613} & 1.096 & 1, 5,6,7 \\
\object{GRB\,980703} & 0.966 & 8 \\
\object{GRB\,990123} & 0.61 &  9 \\
\object{GRB\,990510} & 1.62 & 10, 11 \\
\object{GRB\,991216} & 1.02 & 11, 12 \\
 \noalign{\smallskip}
            \hline
         \end{array}
      \]
\begin{list}{}{}
\item[1] Costa \cite{Costa99}
\item[2] Galama et al. \cite{Gal97}
\item[3] Pedersen et al. \cite{Peder98}
\item[4] Diercks et al. \cite{Diercks98}
\item[5] Djogovsky et al. \cite{Djo98}
\item[6] Hjort et al. \cite{Hjort98}
\item[7] Halpern et al. \cite{Halp98}
\item[8] Vreeswijk et al. \cite{Vree99}
\item[9] Galama et al. \cite{Gal99}
\item[10] Staneck et al. \cite{Sta99}
\item[11] Piro L. \cite{Piro00}
\item[12] Halpern et al. \cite{Halp00}
\end{list}
   \end{table}

   We insist on the fact that we selected GRBs on the basis
   of the consistency of X-ray and optical data. This imply to use
   only the BeppoSAX data for the 2-10 keV band and
   optical magnitudes in the R band.

\section{Normalization of the X-ray data}

   In order to be able to compare the flux in the 2-10keV
   band, we normalized the data to a common distance corresponding to a
   redshift of 1. We applied
the following corrections: distance-luminosity (relative to the
target distance), normalization to the 2-10keV band in flux, and
time dilation of the temporal scale. These corrections have been
computed using the spectral index provided for the GRB source
spectra when available. When the burst X-ray spectrum was not
known with enough accuracy, we used the value of 1, which seems to
be consistent with the majority of the burst afterglow spectra
(Costa \cite{Costa99}). We took as a base a flat universe with
${\Omega}_{\Lambda} = 0.7$. Table 2 details the resulting
correction factors applied to the data.

\begin{table}
\caption[]{X-ray correction factors}
\label{Xray}
 \[
 \begin{array}{llll}

 \noalign{\smallskip}
 \smallskip
            & &\multicolumn{2}{c}{Correction\:factors}\\
            Source&Redshift &Distance&flux\\
 \noalign{\smallskip}
 \hline
 \noalign{\smallskip}
\object{GRB\,970228}& 0.695& 0.41& 0.85\\
\object{GRB\,970508}& 0.835& 0.64 & 0.92\\
\object{GRB\,971214}& 3.42& 20.2& 2.21\\
\object{GRB\,980613}& 1.096& 1.25 & 1.05\\
\object{GRB\,980703}& 0.966& 0.92& 0.99\\
\object{GRB\,990123}& 1.61& 3.23 & 1.19\\
\object{GRB\,990510}& 1.62& 3.29 & 1.31\\
\object{GRB\,991216}& 1.02& 1.05 & 1.01\\
\noalign{\smallskip}
\hline
         \end{array}
\]
\end{table}

Figure 1a displays the X-ray afterglow raw data. The same sample
is displayed figure 1b, but we applied the corrections shown in
table 1 in order to plot the light curves for a standard distance
corresponding to a redshift of 1.

\section{Optical data}

\begin{figure*}
\resizebox{\hsize}{!}{\includegraphics{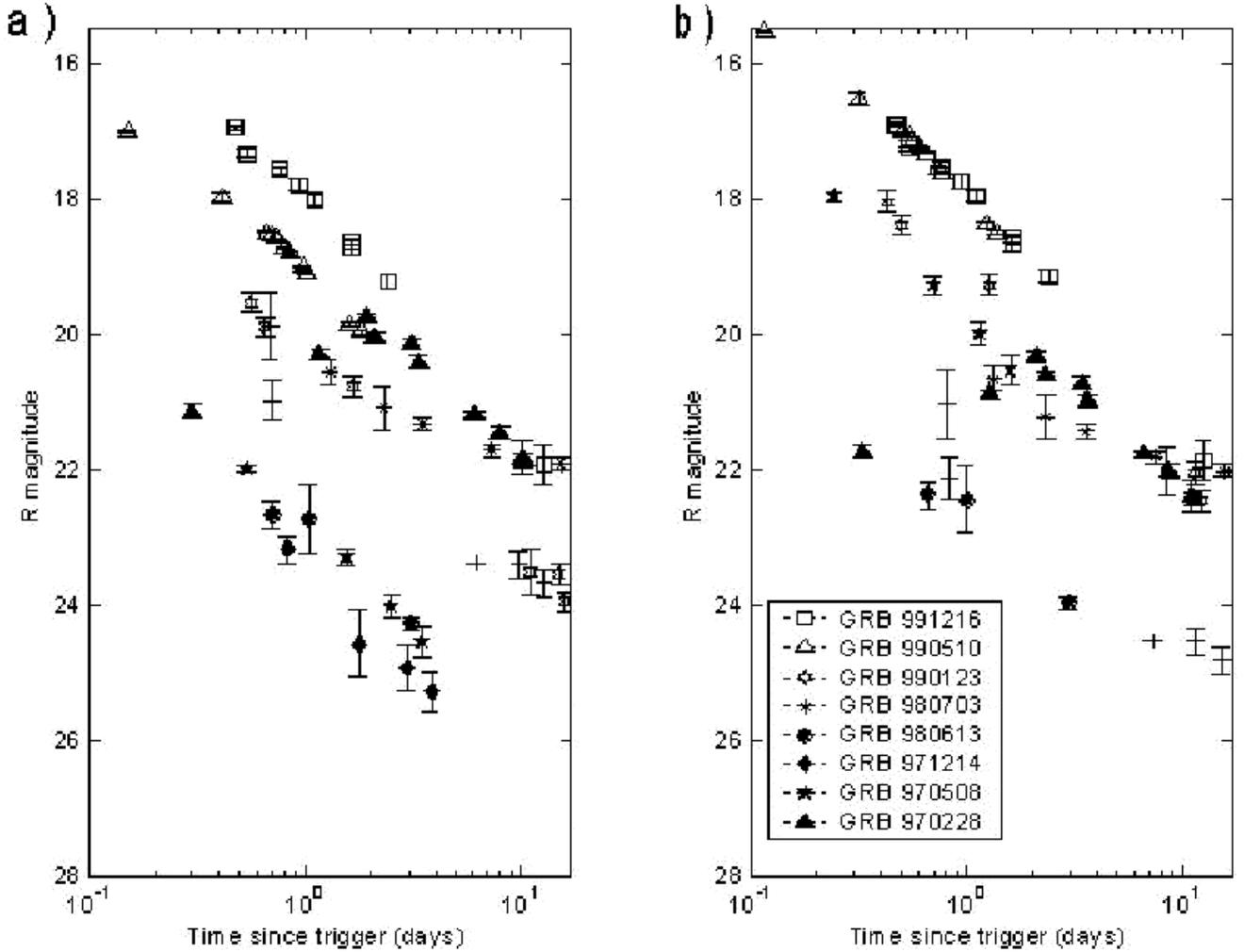}} \caption{Optical
light curves in the R band of the burst sample a) no correction
has been applied, and b) with distance, flux, time dilation, and
R absorption corrections applied, normalizing the distance to a
redshift of 1} \label{OptFig}
\end{figure*}

The same corrections have been computed for the optical data in
the R band from the same sources, of course taking into account
the wavelength. At optical wavelengths the galactic absorption
may be somewhat high. We used the work of Schlegel et al.
(\cite{Schl98}) to derive an additional correction to apply to
the data. The resulting factors are shown on table 3.

\begin{table}
\caption[]{Optical correction factors}
\label{Optics}
 \[
 \begin{array}{llll}
 \noalign{\smallskip}
 \smallskip
            &\multicolumn{3}{c}{Correction\:factors}\\
            Source&Distance&Flux&Absorption\\
 \noalign{\smallskip}
 \hline
 \noalign{\smallskip}
\object{GRB\,970228}& 0.41& 0.85& 0.61\\
\object{GRB\,970508}& 0.64 & 0.92& 0.05\\
\object{GRB\,971214}& 20.2& 1.99& 0.06\\
\object{GRB\,980613}& 1.10 & 1.25& 0.22\\
\object{GRB\,980703}& 0.92& 0.99& 0.72\\
\object{GRB\,990123}& 3.23 & 1.22& 0.04\\
\object{GRB\,990510}& 3.28 & 1.18& 0.53\\
\object{GRB\,991216}& 1.05 & 0.09& 1.67\\
\noalign{\smallskip}
            \hline
         \end{array}
      \]
\end{table}

The raw optical light curves are displayed on figure 2a. On
figure 2b we plotted the corrected light curves, again for a
standard distance corresponding to a redshift of 1. As it can be
seen, the correlation observed at X-ray wavelengths vanishes
almost completely, though the dispersion in magnitude is somewhat
reduced.


\section{Discussion and conclusions}

The X-ray data present, when corrected, an obvious correlation
 into two homogeneous groups. We find a mean slope
 of $-1.6 \pm 0.2$ for the most luminous subset, and
  $-1.11 \pm 0.17$ for the less luminous afterglows.

  If we try to interpret the observed correlation in the framework of
  the fireball model, we have to suppose that the differential
  density of the medium shocked by the fireball
plays a secondary role. Let us refer to the equations (7) to (12)
of Sari at al. (\cite{Sari98}). If we take $\rm{p} = 2.3 \pm 0.15$
for the index of the electron distribution power law, our results
at X-ray wavelengths are compatible with the radiative case for
the most luminous afterglow group, and with the adiabatic case
for the less luminous group. This is also close to the standard
value assumed by Sari et al. (\cite{Sari96}) of $\rm{p} = 2.5$.

However, the afterglow temporal evolution should be dependent on
the surrounding medium density. It is clearly not the case here,
though the burst sample is somewhat restricted. An explanation may
be that the burst source surrounding medium has been "washed"
before the shock. Of course, in the absence of any measure of the
critical transition time of the afterglow light curve, it is
impossible to conclude on the density parameter. However the
correlation we present at X-ray wavelengths seems to be more
compatible with a weak dependence on it, i.e. a somewhat low
density medium. For the GRBs belonging to the most luminous
group, this conclusion is consistent with the results obtained by
Kumar (\cite{Kum2000}) for a sample of 7 bursts in the same 2-10
keV energy range. It is interesting to note that Kumar and Piran
(\cite{Kum2000b}) have shown that the width of the distribution
function for the X-ray afterglow flux should be significantly
smaller that the spread in GRB fluences. The small number of
sources we used in this work prevent from any firm conclusion,
though GRB afterglows in the slow cooling regime may show a
larger dispersion, which may be confirmed with the larger data
set expected from HETE-2 and SWIFT.

It is difficult to find a firm explanation on the absence of
correlation at optical wavelengths. Since the fireball model
predicts that the ratio between the optical and X-ray luminosity
should remain approximately constant during the early phases of
the afterglow, we can try to explain the absence of correlation
by an external reason, such as the absorption in the host galaxy
of the GRB source. In this case the discussion of the above
paragraph does not apply. Assuming that the X-ray and optical
light curves have the same indexes, we computed the absorption
coefficients tabulated in the second column of  table 4.

   All values are relative to the magnitude of GRB 980703.
   For GRB 970508, we computed the correction on the decreasing
   part of the light curve after 1.5 day. This correction is negative for GRB
   970508, the
   most luminous of our sample.
   Since the absorption occurs at UV wavelengths if we
   take into account the source redshift, these values are
   quite reasonable. They correspond to a rough column density of
   $5 \times 10^{21}$ at most (Prehdel and  Schmitt
   \cite{Pred95}). For these values of the host galaxy
   medium column density, the transmission coefficient of 2keV photons is
   above 0.95 (Seward, \cite{Sew2000}).
   In order to assess the role of the UV absorption,
   we computed the decay slopes at X-ray wavelengths,
   and we compare them with the rate of decay of the afterglow light in the R band.
The results are given in the last two columns of table 4.

\begin{table}
\caption[]{R absorption for the optical data and comparison
between the optical and X-ray power law decay indexes.}
\label{Abs}
 \[
 \begin{array}{llll}
 \noalign{\smallskip}
 \smallskip
  &&\multicolumn{2}{c}{Power\:law\:decay\:indexes}\\
 Source& R\:Absorption& X-ray & Optical\\
 \noalign{\smallskip}
 \hline
 \noalign{\smallskip}
\object{GRB\,970228}& 2.0  & 1.28 \pm 0.11 & 1.10 \pm 0.1\\
\object{GRB\,970508}& -0.6 & 1.02 \pm 0.21 & 1.17 \pm 0.04 \\
\object{GRB\,971214}& 2.0  & 1.62 \pm 0.27 & 1.20 \pm 0.02\\
\object{GRB\,980613}& 2.6  & 1.05 \pm 0.37 & 1.00 \pm 0.01\\
\object{GRB\,980703}& 0.0  & 0.57 \pm 0.14 & 1.17 \pm 0.25\\
\object{GRB\,990123}& 2.3  & 1.44 \pm 0.20 & 1.44 \pm 0.07 \\
\object{GRB\,990510}& 0.5  & 1.54 \pm 0.15 & 0.82 \pm 0.02\\
\object{GRB\,991216}& 1.1  & 1.78 \pm 0.01 & 1.22 \pm 0.04\\
\noalign{\smallskip}
            \hline
         \end{array}
      \]
\end{table}

The mean power law decay index at X-ray wavelengths is $1.6 \pm
0.2$ for the most luminous afterglow group, and $1.11 \pm 0.17$
for the less luminous one. At optical wavelengths the values are
respectively $ 1.17 \pm 0.13$, and $1.09 \pm 0.07$. Though there
is on average some difference at optical wavelengths also between
the two groups, this feature cannot be considered as established.

We have presented in this letter a correlation between the
afterglow light curves at X-ray wavelengths. While this behaviour
is not observed in the optical domain, the decreasing rate may
appear to be correlated in both energy ranges. The fireball model
provides a framework to explain the X-ray light curves, while the
host galactic absorption may shade the correlation in the visible
range. The HETE-2 experiment, due to launch this year, will
provide new data which may confirm this picture.

\begin{acknowledgements}
We thank J.L. Att\'eia, R. Mochkovitch, P. M\'esz\'aros and the
anonymous referee for helpful comments. Thanks are also due to J.
Greiner who maintains a web page
(http://www.aip.de/$\sim$jcg/grb.html) with exhaustive data on
burst afterglows.

\end{acknowledgements}

\end{document}